\documentclass{webofc}
\usepackage[varg]{txfonts}   
\usepackage[bf]{caption}
\usepackage{subcaption}
\usepackage{graphicx}
\usepackage{listings}
\usepackage{hyperref}
\usepackage{amsmath}
\usepackage{siunitx}
\newcommand{\diff}{{\rm d}} 

\newcommand{\CR}{C_{\rm R}^{}}

\newcommand{\ptrel}{\boldsymbol{p}_{\rm rel}}

\newcommand{\invtf}{t^{-1}_{\rm form}}

\newcommand{\mycomment}[1]{} 

\begin{document}
\title{Parton cascades at DLA: the role of the evolution variable}
%
%

\author{%
\firstname{Carlota} \lastname{Andrés}\inst{1}
\and
\firstname{Liliana} \lastname{Apolinário}\inst{2,3}
\and
\firstname{Néstor} \lastname{Armesto}\inst{4}
\and
\firstname{André} \lastname{Cordeiro}\inst{2,3}\fnsep\thanks{\email{andre.cordeiro@tecnico.ulisboa.pt}}
\and
\firstname{Fabio} \lastname{Dominguez}\inst{4}
\and
\firstname{José Guilherme} \lastname{Milhano}\inst{2,3}
}


\institute{CPHT, CNRS, École polytechnique, Institut Polytechnique de Paris \\ 91120 Palaiseau, France
\and
           Laboratório de Instrumentação e Física Experimental de Partículas (LIP)  \\ Avenida Professor Gama Pinto, 2, 1649-003 Lisboa, Portugal
\and
           Departamento de Física,  Instituto Superior Técnico, Universidade de Lisboa \\ Avenida Rovisco Pais 1, 1049-001 Lisboa, Portugal
\and
           Instituto Galego de Física de Altas Enerxías (IGFAE), Universidade de Santiago de Compostela \\ Santiago de Compostela 15782, Spain 
          }





\abstract{%
The theoretical treatment of jet quenching lacks a full description of the interplay between vacuum-like emissions, usually formulated in momentum space, and medium induced ones that demand an interface with a space-time picture of the expanding medium and thus must be formulated in position space. 
%
%
In this work we build a toy Monte-Carlo parton shower ordered in formation time, virtual mass, and opening angle, which are equivalent at leading logarithmic accuracy. Aiming to explore a link with jet substructure, we compute the Lund plane distributions for the different ordering prescriptions. Further, we investigate the sensitivity of ordering prescriptions to medium effects by counting the number of events eliminated by a decoherence condition.
}
\maketitle
\vspace{-0.5cm}
\section{Introduction}
\label{intro}

Ultra-relativistic heavy ion colliders, such as the Large Hadron Collider (LHC) and the Relativistic Heavy Ion Collider (RHIC) have unlocked the study of the Quark-Gluon Plasma (QGP), a hot and dense nuclear medium characterised by color deconfinement.
Studies of this state of matter are made possible by the analysis of high-energy jets produced in the collision. Compared to proton-proton collisions, their energy and substructure are modified through interactions with the medium~\cite{Apolinario:2022vzg}, and the wide range of energy scales present within a jet allows for the construction of jet quenching observables which might unlock a space-time tomography of the QGP~\cite{Apolinario:2017sob,Apolinario:2020uvt,Andres:2016iys}. However, a theoretical description of jet-medium interactions requires accounting for the interplay between vacuum-like and medium-induced emissions~\cite{Caucal:2018dla}, complicated by the fact that the former are formulated in momentum space while the latter require an interface with a space-time picture of an evolving medium.

Aiming at such a description, this work explores three different formulations of a vacuum-cascade at double-logarithmic accuracy, based on the resummed no-emission probability computed from perturbative QCD, ordered in formation time, invariant mass, and opening angle~\cite{Sjostrand:2006za,Corcella:2000bw}. The ambiguity inherent to this choice of ordering variable is reflected in the shower kinematics \cite{Nagy:2014nqa} and can be understood as characterising the uncertainty of this approximation. We further explore the impact of these different formulations on jet-medium interactions by implementing a simplified jet quenching model.

\section{Building differently ordered cascades}
\label{sec:1}

The scale evolution of a QCD cascade is generated by sampling the \textit{no-emission probability} between a previous scale $s_{\rm prev}$ and the next scale $s$ , given by the Sudakov form factor,
\begin{align}
    \Delta_{ R}(s_{\rm prev}, s_{}) = \exp\left\{ 
    - \, \frac{\alpha\,\CR}{\pi}
    \int^{s}_{s_{\rm prev}} \frac{\diff s}{s} 
    \int^{}_{\Gamma(s)} \frac{\diff z}{z} 
    \right\}
    \,,
\label{eq:no-emission-probability}
\end{align}
\noindent obtained by resumming the differential rate for resolvable emissions calculable in perturbative QCD. This requires integrating over all possible emission scales $s$ and splitting fractions $z$ and taking the leading logarithmic contribution for the splitting function, $\hat{P}(z) \simeq 2 C_{R} / z$, where $R$ stands for the color representation of the emitter.
The integration range in \eqref{eq:no-emission-probability} is due to the resolution criterion for the splittings, corresponding to a transverse momentum cutoff. 

We use light-cone coordinates, in which the four-momenta are written $p^\mu=(p^+, p^-, \boldsymbol{p})$, such that $p^\pm = (p^0 \pm p^3)/ \sqrt{2}$, and $\boldsymbol{p}$ stands for the transverse momentum components.
In these coordinates, the kinematic variables for a generic splitting $p^{}_a \to p^{}_{b} + p^{}_{c}$ are written
%
%
$\ptrel = (1-z)\,\boldsymbol{p}_b - z\,\boldsymbol{p}_c$ and $z = p^+_b / p^+_a = 1 - p^+_c / p^+_a$ where $\ptrel$ stands for the relative transverse momentum of the daughter partons, and $z$ is the light-cone momentum fraction of parton $b$. Based on these definitions, we choose three different ordering variables, namely the inverse formation time $\invtf$, the invariant mass $p^2$, and the (squared) opening angle $\zeta$, defined respectively as
\begin{align}
    \invtf  &= \frac{|\ptrel|^2}{E \, z(1-z)}       \,,\quad
    p^2     = \frac{|\ptrel|^2}{z(1-z)}            \,,\quad
    \zeta   = \frac{|\ptrel|^2}{[E \, z(1-z)]^2}   
    \,,\quad \textrm{ with } \quad
    E = p_{a}^+ / \sqrt{2}
    \,.
\label{eq:ordering-variables-def}
\end{align}
%

We take the hadronisation condition as $|\ptrel|=\Lambda$, allowing the shower to continue while the inequality $|\ptrel|^2 > \Lambda^2$ is verified. This condition can be rewritten in terms of each ordering variable, providing a regularisation of the soft divergence. In the case of formation time ordering, for example, this condition is written
\begin{align}
    z(1-z) &> \frac{\Lambda^2 / E}{\invtf} 
    \,.
    \label{eq:z-phasespace}
\end{align}
%
%
Because the left-hand side of this inequality has an upper bound of $1/4 > z(1-z)$, this condition also provides the lower bound on the formation time, which we identify as the hadronisation time $\invtf > 4\Lambda^2 / E$.
The starting condition of the shower is set by requiring an upper bound on one of the ordering variables. A natural choice is that the formation time of the first splitting must be larger than the time-scale of the hard scattering, set by the energy of the first quark, $\invtf < E$. We further impose an the angular restriction $\zeta < 4$ in order to ensure the consistency of the $\zeta$ distribution (among others) across ordering prescriptions.

The procedure to generate a parton cascade can be illustrated by considering formation time ordered showers. One begins by generating a value for $\invtf$ from the no-emission probability in \eqref{eq:no-emission-probability}, along with a light-cone fraction $z$ from the splitting function $\hat{P}(z) \propto 1/z$. From the $(\invtf, z)$ pair, we compute the angular variable $\zeta$ and reject all splittings which do not obey $\zeta < 4$, continuing the sampling procedure. Parton splittings continue to be generated until the inverse formation time reaches its lower bound given by the hadronisation condition.

In the remainder of this work we describe preliminary results for $10^6$ parton cascades sampled according to each of the three ordering prescriptions, initiated by a quark with light-cone energy of $E_{\rm jet} = \SI{1}{GeV}$ and a hadronisation cutoff of $\Lambda = \SI{1}{GeV/c}$.

\vspace{-0.30cm}

\section{Lund Plane densities and trajectories}
\label{sec:3}

\begin{figure}[h]
\centering
\begin{subfigure}{.435\textwidth}
    \centering
    \includegraphics[width=\textwidth]{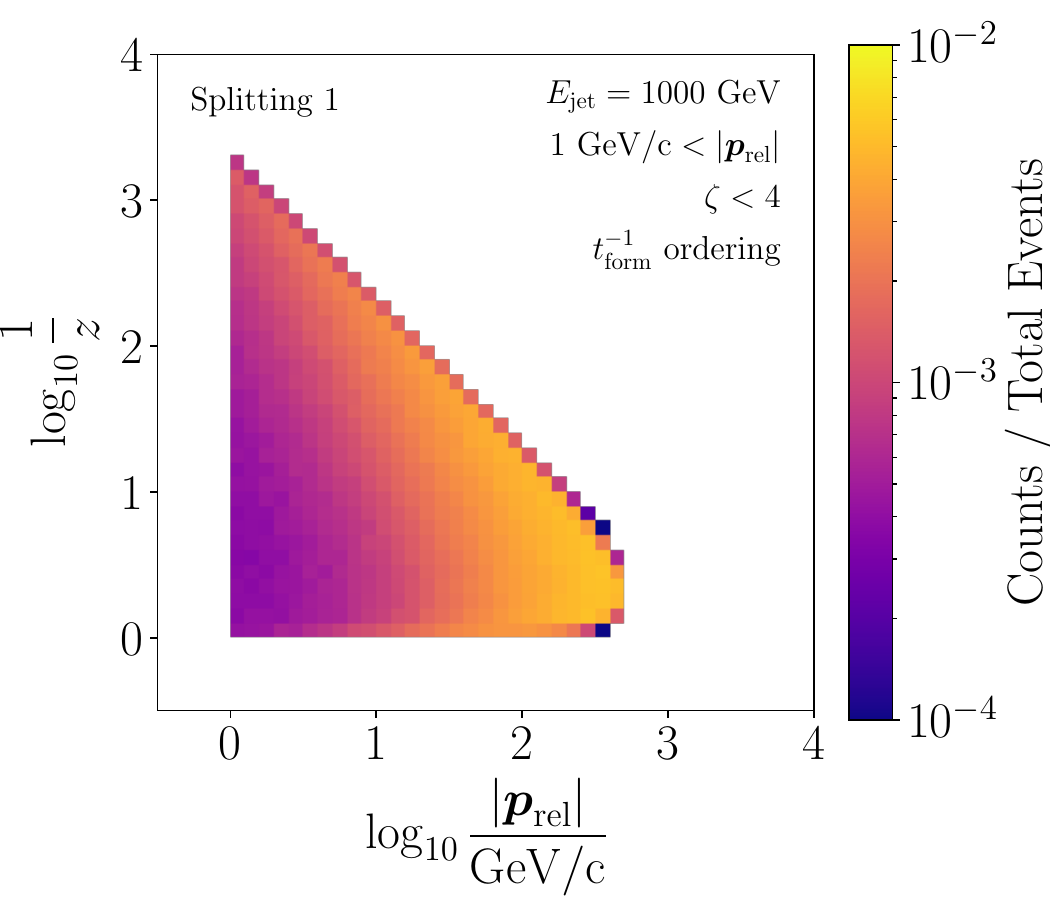}
    \label{subfig:lund-plane}
\end{subfigure}%
\begin{subfigure}{.375\textwidth}
    \centering
    \includegraphics[width=\textwidth]{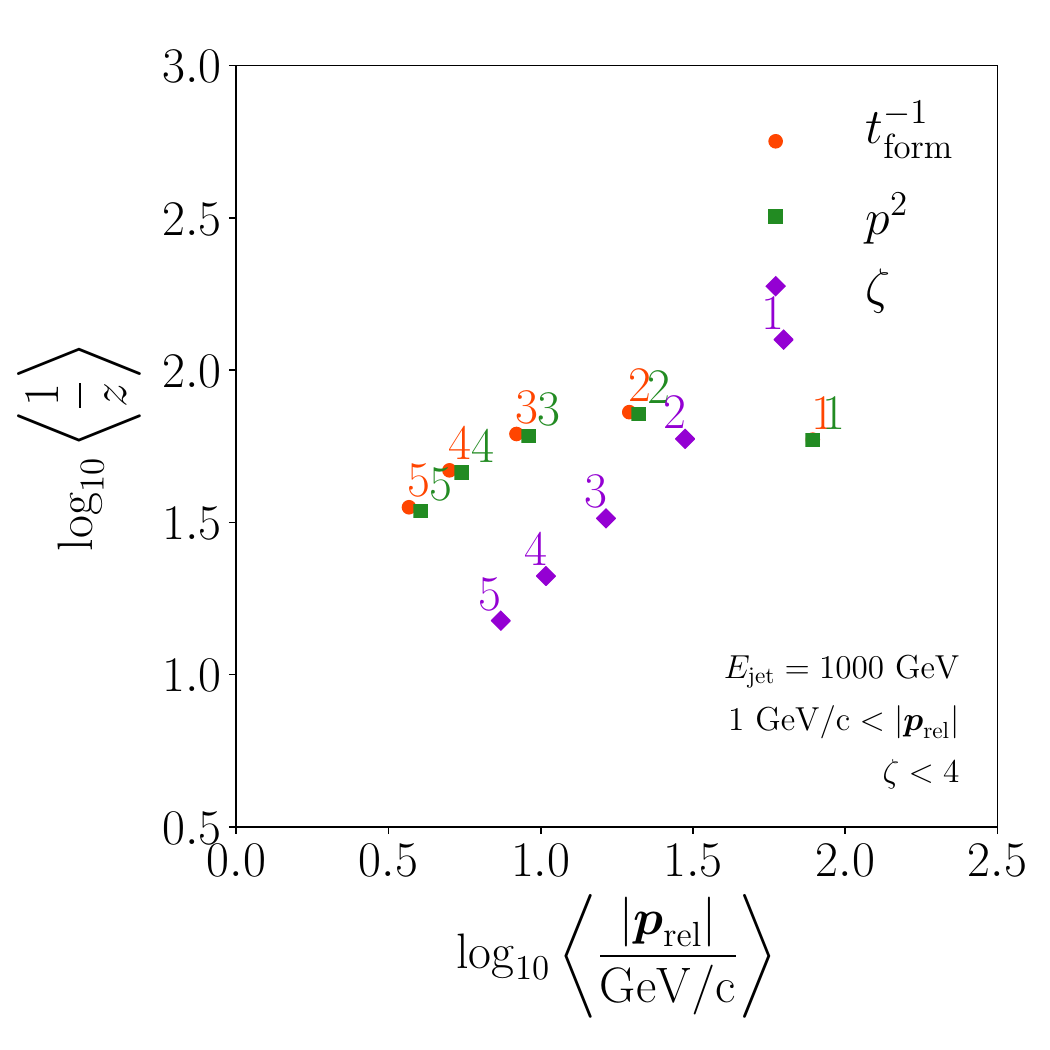}
    \label{subfig:lund-traj}
\end{subfigure}
\vspace{-0.5cm}
\caption{\textbf{(Left)} First splitting Lund distribution for cascades ordered in formation time. \textbf{(Right)} Lund trajectories for cascades ordered in formation time (orange), virtual mass (green), and angle (purple).}
\vspace{-0.3cm}
\label{fig:lund-plots}
\end{figure}

Lund diagrams \cite{Dreyer:2018nbf} allow for a representation of of parton emissions as points in a plane with coordinates $\big\{ \log_{10}(|\ptrel| / \SI{}{GeV/c}), \log_{10}(1/z) \big\}$.
Focusing on the first splitting in the parton cascades, we represent the primary Lund plane in figure~\ref{fig:lund-plots}~(left) for time ordered cascades, noting how large transverse momenta and small light cone fractions are favoured. A comparison between ordering prescriptions is enabled by considering the ``trajectories'' of the quark splittings in this Lund plane, i.e the mean values of $|\ptrel|$ and $z$ distributions for the first five splittings along the quark branch. This is shown in figure~\ref{fig:lund-plots}~(right) for all three ordering prescriptions. We find an evolution towards increasing values of the light-cone fraction (relative to the quark), consistend with the closing of phase-space predicted by equation~\eqref{eq:z-phasespace}, although the specific trajectories vary with the ordering prescription.
%
%
These results imply that differently ordered cascades can differ significantly in their kinematics, resulting in a different distribution of formation times, which may play a determining role in coupling the shower evolution with a finite medium. In order to explore this possibility the next section presents a simplified quenching model along with its impact on differently ordered cascades.

\vspace{-0.15cm}

\vspace{-0.90cm}
\section{A simple quenching model}
\label{sec:4}

We implement a simple quenching model by
reducing the parton-medium interactions to a simple condition, stating that a cascade is suppressed if, in any of its quark-initiated splittings the relative transverse distance acquired by the daughters exceeds the inverse saturation scale of the medium, $Q^{-1}_{\rm sat} = (\hat{q} L)^{-1/2}$, where the transport coefficient $\hat{q}$ depends on the medium density and $L$ is the medium length. We further demand that only splittings inside the medium can be modified, $t_{\rm form} < L$. 
These conditions are encoded in the quenching probability
%
\begin{align}
    \mathcal{P}_{\rm quenching} = 
    \Theta\left( L - t_{\rm form}\right)
    \times
    \Theta\left( r_{\rm split } - (\hat{q} L)^{-1/2} \right)
    \,,\quad
    \textrm{ where }
    \quad
    r_{\rm split } = \frac{1}{|\ptrel|}
    \,.
\label{eq:quenching-model}
\end{align}

\begin{figure}[h]
\centering
\includegraphics[width=0.50\textwidth]{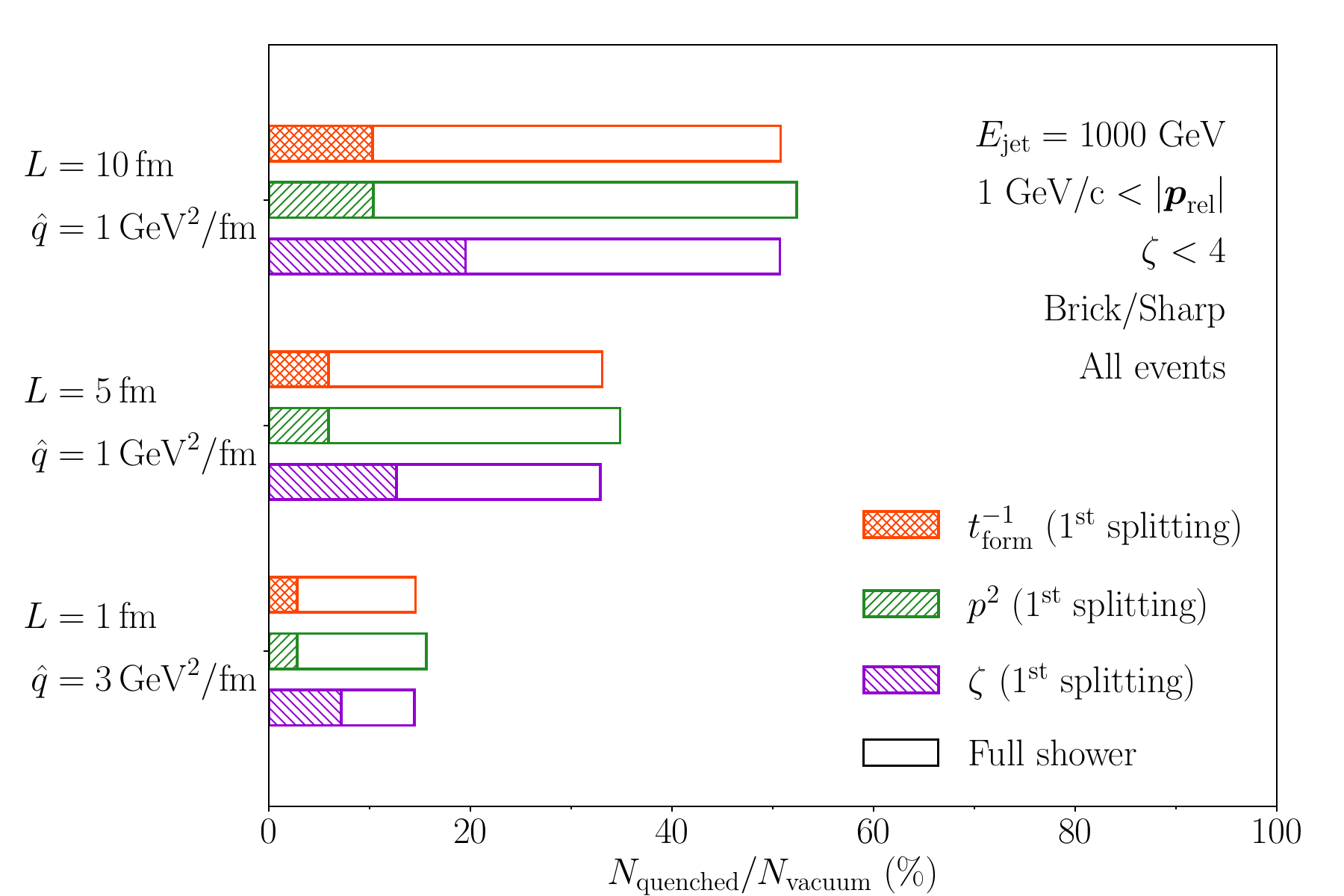}
\caption{Percentage of quenched events for the three ordering variables, when the condition is applied to the first splitting (full bars) or to the full quark branch (empty bars) for different sets of quenching parameters.}
\label{fig:quenching-weights}
\vspace{-0.7cm}
\end{figure}

We apply this model to cascades ordered according to each of the prescriptions, in two different modes; either checking the conditions in \eqref{eq:quenching-model} only against against the first emission of the cascade, or against all splittings in the quark branch. The percentage of eliminated events for both cases is displayed in figure~\ref{fig:quenching-weights} for three different sets of quenching parameters $(\hat{q}, L)$. While we find significant differences between algorithms when applying the condition to the first splitting, these diminish when the full quark branch is taken into account.
These results suggest that while the integrated energy loss of a parton cascade is not sensitive to the ordering prescription at DLA accuracy, medium-induced effects over the first few splittings might be affected by this choice. Compounded by lack of medium evolution in this simplified model, these findings motivate the need for a theoretically consistent formalism to describe the interface between a developing jet and an evolving medium.

\vspace{-1.7cm}
\section{Summary}
\label{sec:5}

In this work we developed a toy Monte Carlo for parton showers at double logarithmic accuracy. The setup allows to consistently change between different ordering variables that were chosen to be formation time, virtual mass and opening angle.
The Lund plane densities and trajectories for all three prescriptions were computed. Moreover, we applied a simplified quenching model inspired by coherence effects in a finite size medium of constant density, obtaining the percentage of suppressed events for different quenching scenarios. While this quenching frequency seems independent of the the ordering prescription when the model is applied to the full shower, significant differences remain when applying the model exclusively to the first emission, highlighting the need for an interface between jet development and medium evolution.


\textbf{Acknowledgements}
This work has received funding by OE Portugal, Fundação para a Ciência e a Tecnologia (FCT), I.P., projects EXPL/FIS-PAR/0905/2021 and CERN/FIS-PAR/0032/2021; by European Research Council project ERC-2018-ADG-835105 YoctoLHC and has received funding from the European Union’s Horizon 2020 research and innovation programme under grant agreement No.824093. 
A.C. was directly supported by FCT under contract PRT/BD/154190/2022.
%
%

\vspace{-0.40cm}

\bibliography{bibliography.bib}

\end{document}